# Building Faculty Expertise Ontology using Protégé: Enhancing Academic Library Research Services

**Snehasish Paul**

*Department of Library and Information Science, University of Delhi, Delhi- 110007, India*
*E-mail: snehasishpaulas98@gmail.com*
*ORCID- https://orcid.org/0009-0003-2730-5314*

**ABSTRACT**

Academic libraries struggle to find and access faculty expertise across disciplines. This research proposes a faculty expertise ontology with a hierarchical structure based on Protégé to enhance library services and knowledge organisation. The ontology classifies relationships between departments, subject areas, faculty members, and contact data into layers including Top, Middle, and Bottom levels. The academic structure that this tiered form takes enables discovery of expertise in departments. The ontology which answers competency questions generated from the subject matter experts can answer real-world questions like which faculties are in the specific areas, how to collaborate with other disciplines and search contact information and so on. Competency questions act as design and test instruments to show that the ontology will fulfil the information needs of Researchers, Librarians and Administrators. The ontology is able to cope with semantically-enhanced queries, as shown by SPARQL implementations. The model works effectively in initiating referrals to an expert, aligning research with the strength of a department and allowing academics to partner up. The ontology delivers a scalable platform that adapts to institutional change. In the future, we intend to integrate with institutional databases and library systems for automatic API updates, as well as develop user interfaces and visualisations.

## 1. INTRODUCTION

Academic libraries are key institutional knowledge centres in universities that bridge experts with specialised information. However, it is challenging for librarians and experts to efficiently identify subject matter experts across multiple disciplines. This paper proposes a novel solution to this challenge through the construction of faculty expertise ontologies using the well-known open-source ontology editor Protégé (Noy & Mcguinness, 2001).
Our ontology scheme hierarchically structures faculty expertise in terms of academic disciplines, from departments to research-specialist areas. By quantifying the relationships between faculty members, their departments, subject specialisations, and their contact details, this knowledge graph facilitates precise queries of institutional expertise. Multiple paths to finding relevant faculty experts are supported by a top-level, middle-level, and bottom-level hierarchical classification scheme (*View of Ontology Design Of A Modern Learning Environment And Modern Pedagogy Using Protégé Software*, 2024).
This ontological strategy has great advantages for academic library research services, enabling purpose-specific collaborations, multidisciplinary relationships, and enhanced resource management. Librarians can use this strategy to facilitate a match between researchers and relevant faculty subject experts, improve collection development based on institutional strengths, and assist in strategic studies. With universities focusing more on knowledge sharing and cross-disciplinary research collaboration, systematic expertise mapping is a powerful instrument for leveraging institutional intellectual capital (Utkucu & Sacks, 2025a). To support this objective, an ontology of faculty expertise was built in Protégé using a three-level hierarchical classification model of the Top, Middle, and Bottom. This multilayered structure reflects the departmental and programmatic flow of academia and nuanced specialist areas within the faculty domain. The resultant ontology presented through visual models shows how subject areas, departmental and disciplinary affiliations, and individual expertise are related in an organised and structured manner. These relationships serve as the foundation upon which exacting knowledge can be retrieved and the basis upon which to solve specific information requests, as measured through a set of competency questions developed with particular care (Varadarajan et al., 2023).

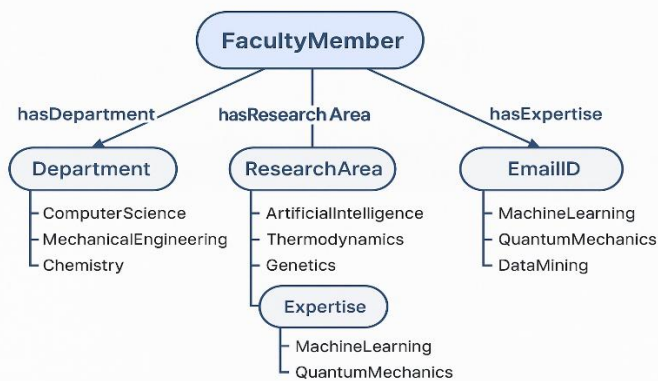

**Source(s):** created by author
**Figure 1: Flow chart representation of the Faculty Member Ontology**

This flowchart illustrates the connections between the Faculty Member concept and various academic departments, research fields, and areas of expertise, offering a clear depiction of the ontology.

## 2. OBJECTIVES OF THE STUDY

1. To model faculty expertise hierarchically using Protégé.
2. To classify academic domains into Top, Middle, and Bottom levels.
3. To define semantic links between faculty, subjects, and contact details.
4. To support expert identification for research and collaboration.
5. To ensure the scalability and adaptability of the ontology structure.

## 3. LITERATURE REVIEW

Although several ontologies in fields such as food, culture, and energy already exist, no in-depth ontology models are specifically tailored to represent faculty expertise in research-support services. This emphasises the necessity of domain-specific ontologies for academic knowledge management and retrieval.

Underscored Protégé's value in streamlining ontology development with its easy-to-use interface, strong support for class hierarchies, and the visualisation of semantic relations. It also supports semantic clarity, interoperability, and information reuse. However, the research was within the domain of energy access (Sharma & Kanjilal, 2023). Examined the use of ontology rules to support research services at the Khon Kaen University Library. Using Protégé version 3.5 and the Ontology Web Language (OWL), this study created a ten-specific rule-based knowledge base covering plagiarism detection, journal assessments, information retrieval, and training programs. The ontology performed better in terms of performance metrics, with a precision of 92.73%, recall of 90.74%, and F-measure of 91.72%, proving its effectiveness in supporting academic research. Although this research did not specifically address the field of faculty expertise ontologies, its approach to the development and access to structured systems for knowledge representation provides valuable contributions (Ruger et al., 2022). Created an ontology to conserve the cultural knowledge base of traditional dance, focusing specifically on the Rabha tribe of Northeast India. They used Protégé and a facet-based classification scheme to model the rich sociocultural aspects of traditional dance, including costumes, occasions, instruments, and dance movements. The ontology was evaluated using expert-based competency questions and SPARQL queries. The resulting implications of their work highlight what ontology can do for cultural preservation, useful where smart information retrieval is required, and as a component in e-learning and digital heritage environments (Kalita & Deka, 2020). Proposed the YAMO methodology for large-scale ontology building, exemplified by a food-domain ontology. This methodology uses top-down and bottom-up approaches, yielding a flexible, scalable, and comprehensive ontology. The system could respond correctly to 94% of the user requests, proving its efficacy. YAMO provides a structured generic template that can be used in many domains for systematic ontology development (Dutta et al., 2015).

Investigated how the OWL API could be used with Protégé for effective ontology parsing, considering the loading, extraction, or manipulation of ontology constructs such as classes, properties, and relations. This paper does not specifically deal with faculty specialty ontologies or academic library services but is a good source of insight into ontology structure management as well as parsing strategies. This study emphasises the superiority of the OWL API compared to the older Jena system, especially its applicability in terms of scalable and effective ontology management. These approaches can be used in the technical design of academic ontologies, such as those for mapping faculty expertise (Zhao et al., 2012).

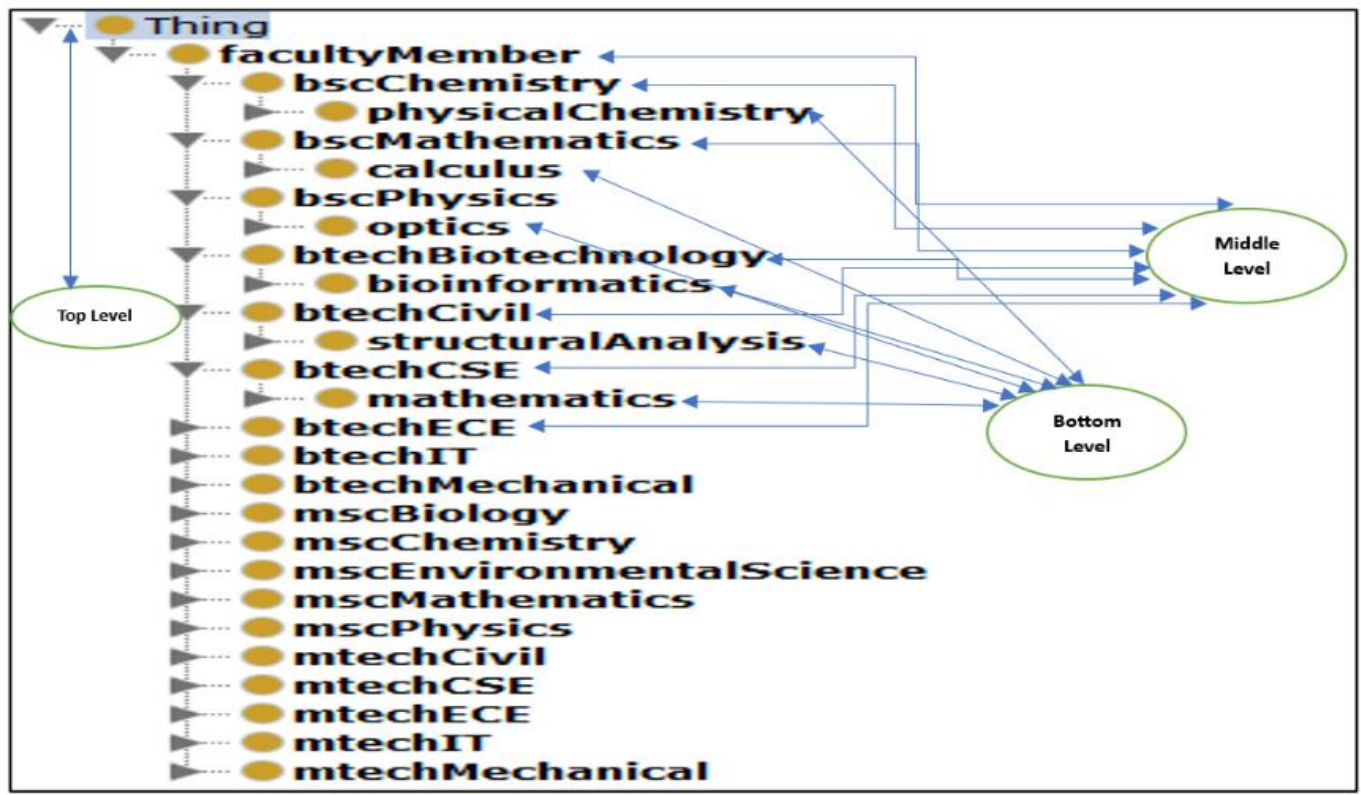

Figure 2: Hierarchical organization of faculty expertise ontology (Top-Middle-Bottom Levels)

## 4. RESEARCH METHODOLOGY

The approach followed in this study is a systematic ontology engineering design, development, and validation of a hierarchical faculty expertise ontology with the aid of Protégé. This procedure comprises ontology planning, class hierarchy design, specification of properties, population, and evaluation using competency questions. It is based on theoretical concepts and practical academic library research service requirements

### 4.1 Ontology development tool: Protégé

Protégé, an open-source ontology editor developed at Stanford University, was used in this project based on its adaptability, ease of use, and availability of support for OWL. This allows complex ontologies to be built with support for reasoning capabilities in the consistency and queries were checked. It also supports class hierarchy visualisation, which is vital for representing structures to non-technical stakeholders (Clarkson et al., 2025).

### 4.2 Ontology Design Process

Domain analysis began by gathering information from academic institutions, including university homepages, departmental course lists and faculty directories. A three-tiered hierarchical organisation was envisioned as follows:

Top-Level Classes are broad university entities, such as faculties, departments, and academic programs.

Middle-Level Classes encompass individual disciplines or fields of study, such as Computer Science, Mechanical Engineering, or Biotechnology.

Bottom-Level Classes are fine-grained specialties or areas of expertise like Data Mining, Fluid Mechanics, or Plant Genetics.

### 4.3 Class and Property Definitions

Every class was specified with explicit definitions, and object and data properties were established to describe the relationships between them. Some examples are:

1. **hasFacultyMember:** links departments to individual faculty.

2. **hasExpertiseIn:** Connects a faculty member to their subject area.

3. **hasEmail and hasName:** Data properties for contact information.

Domain and range constraints were enforced for logical consistency. Cardinality restrictions were included wherever required to represent one-to-many or one-to-one relationships.(Vigo et al., 2019)

### 4.4 Competency Questions

A series of competency questions were established during the design stage to inform ontology structure and functional coverage. These questions describe what the ontology can output an answer to, such as *"Which faculty members specialize in Data Science?" or "Who are the experts in interdisciplinary fields like Bioinformatics?"* Both the design and validation stages employed these questions to ensure that ontologies facilitated significant queries employing SPARQL or DL queries in Protégé.(Milosz et al., 2024)

### 4.5 Ontology Population

To ensure practical usage simulation, dummy information was generated and used to populate the ontology. Although fictional, the information was formatted to represent the academic structure of an actual university, with departments, programs, and faculty positions. The courses were populated with illustrative faculty names, departments, fields of study, and contact information to represent realistic academic situations. Utilising this process ensured that the ontology model conformed to a real university structure without the actual usage of data. Object properties were used to define the relationships between entities, such as faculty teaching a course, belonging to a department, and maintaining certain research interests. The populated ontology shows that academic library services can capitalise on structured expressions of institutional knowledge to facilitate expert discovery, collaboration, and research facilitation.

### 4.6 Validation and Refinement

Competency questions were tested to assess the ability of the ontology to obtain accurate and relevant results. Structural improvements were made based on the gaps identified in the process.

### 4.7 Alignment with Ranganathan's Principles

The ontology structure complies with S. R. Ranganathan's library classification principles, especially his faceted method of knowledge organisation. The setting of well-defined classes, such as Faculty, Department, Course, and Field of Study, is reflected in Ranganathan's concept of compartmentalisation of knowledge. While his Colon Classification system facilitated multidimensional access to knowledge, ontology facilitated flexible and accurate searches through semantic relationships. It also complies with his fifth law, ***"The library is a growing organism"*** by supporting ongoing amendments and extensions. Through this, ontology updates the classic classification for the digital academic setting. The convergence of the two enhances the universality of Ranganathan's vision for modern-day knowledge systems (*Colon Classification : S. R. Ranganathan :Free Download, Borrow, and Streaming : Internet Archive*, n.d.).

### 5. ONTOLOGY STRUCTURE AND SEMANTIC RELATIONS

The ontology developed to map faculty expertise was designed to align with an academic university hierarchy to be easily understandable, discoverable, and semantically rich. It is a three-level, class-based structure with semantic relationships that connect important academic elements, such as departments, programs, specialisations, and faculty (Antoniou et al., 2024).

### 5.1 Class Hierarchy

The central structure of the ontology is based on the following levels:

**Top-Level Classes:** These are high-level institutional categories, such as Faculty, Department, and Academic Program.

**Middle-Level Classes:** These identify subject fields or disciplines (for example, Computer Science, Mechanical Engineering, Mathematics), generally corresponding to study programs such as B.Tech, M.Tech, B.Sc, and M.Sc.

**Bottom-Level Classes:** These individual areas of research or instructional expertise (e.g. artificial intelligence,

thermodynamics, genetics, and nanotechnology), usually corresponding to faculty members' areas of research. Each class contained suitable annotations (labels and comments) to support readability and reuse.

## 5.2 Object Properties (Semantic Relationships)

To facilitate rich semantic connectivity, several object properties were established with designated domains and ranges. The properties of these materials are as follows:

**belongsToDepartment:** Connects a faculty member to their department.

**assignsTeachesIn:** Links a FacultyMember to an AcademicProgram

**hasExpertiseIn:** Links a faculty member to a subject area or specialisation.

**collaboratesWith:** Illustrates intra-departmental or interdisciplinary collaborations among the faculty members.

These properties facilitate institutional and subject boundary traversal and queries. These qualities enable easy information retrieval by users and are critical for professional referral services.

## 5.3 Inference and Argument

The design aimed to enable logical inference. For example:

1. If a faculty member has expertise in Artificial Intelligence and Artificial Intelligence is a subclass of Computer Science, then we can conclude that the faculty member belongs to the broader Computer Science domain.
2. Patterns of collaboration can be inferred from mutual specialisations or shared teaching loads among programs.

Inferences such as these enable librarians and researchers to identify indirect or latent relationships that enhance the applicability of ontology to real-world settings.

## 5.4 Visualization of Class Structure

Visual diagrams of the class hierarchy and their interrelations were constructed using Protégé tools to create ontology graphs. These diagrams are included in the Appendix and are mentioned in the main text to represent the lucidity and sophistication of the structures. The hierarchy lists Things as the top class, with a branching structure indicating departments, subject areas, and, ultimately, individual expertise.

## 5.5 Use Case: Library Services Expert Referral

**Problem:**
One of our librarians was asked to identify a Subject Matter Expert in Quantum Mechanics for a local community project.

**Ontology solution:**
By taking advantage of the semantics of ontology, librarians can easily map subject areas to faculty members.

**Example Output:**

- **Name:** Priya Sharma
- **Department:** MSc Physics
- **Email:** priyash@university.edu
- **Specialization:** Quantum Mechanics

Figure 3 OntoGraf depicts the semantic connections among significant entities in the Faculty Expertise Ontology. It visually demonstrates how faculty members, such as Priya Sharma, are associated with their respective departments and fields of expertise, like Quantum Mechanics. This organised depiction allows librarians to effectively locate subject matter experts by querying the ontology, thereby facilitating expert discovery for both academic and community-oriented projects.

**Result:**

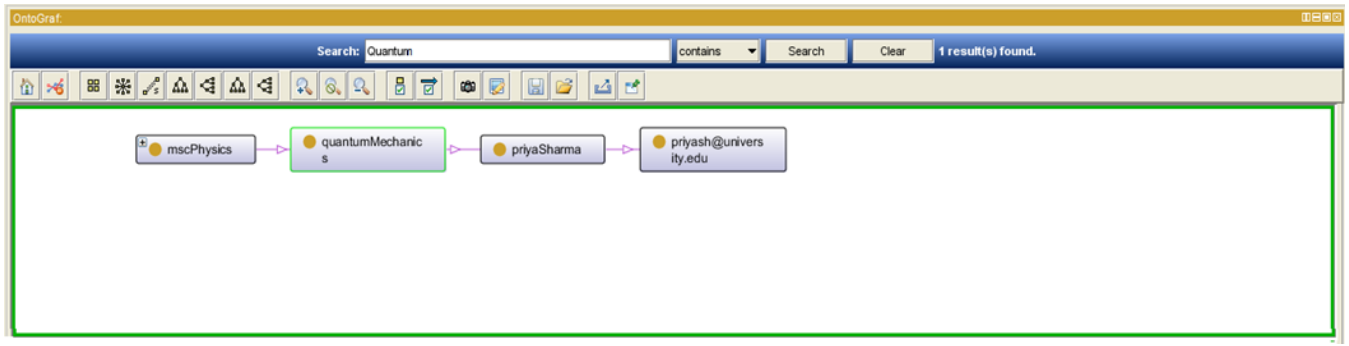

Figure 3: OntoGraf Representation for Case Study – Expert Identification

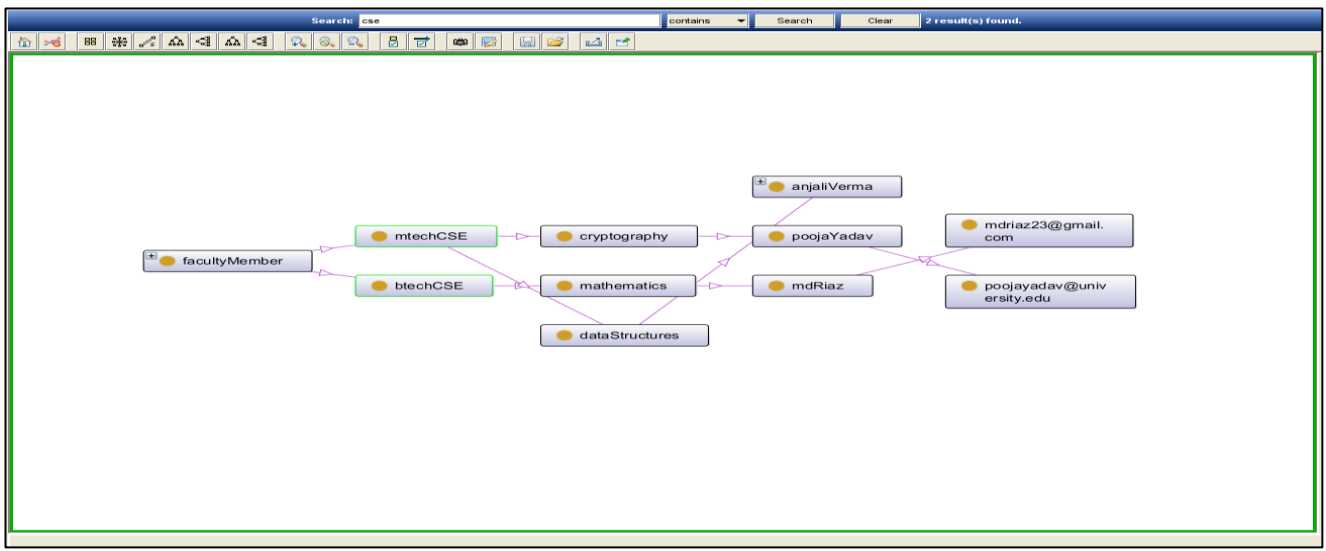

Figure 4: Ontographic visualization in the CSE domain

**Real-world impact:** Empowering libraries to play an active role in facilitating knowledge discovery is essential. These use cases demonstrate the versatility and relevance of ontologies. University staff information is converted into an active, queryable knowledge graph that can support improved decision-making, tighter collaborations, and more adaptive academic services.

### 6. Ontology Evaluation

The ontology was tested using the following competency questions to check for coverage, navigability and semantic consistency. These questions verify that the model achieves its desired purpose of expertise detection, interdisciplinary cooperation, and academic navigation.(Diamantini et al., 2016)

### 6.1 Competency Questions

1. Who are the faculty members in the Mathematics BSc department?
2. Who is responsible for teaching Calculus in the BSc Math program?
3. Which faculties are associated with the MTech CSE program?
4. Which CSE faculty members teach Data Structures?
5. Which departments handle Cryptography?
6. What subjects does Yadav teach?
7. Which faculty members have expertise in Environmental Science?
8. Who among the staff are experts in physics and optics?
9. Which faculties are engaged in both Mathematics and Computer Science?
10. Who can be contacted regarding interdisciplinary studies in Environmental Science and Civil Engineering?

11. Does the department have faculty members with expertise in data science-related research?
12. Who handles applied mathematics relevant to structural analysis?
13. What are the email addresses of all CSE faculty members?
14. Which faculty members have their email addresses listed?
15. Who should be consulted for guidance on Postgraduate Cryptography?
16. What topics are included in BTech CSE?
17. Who provided the data structures course in the Computer Science Department?
18. How is the faculty member Md. Riaz is semantically connected in the model.
19. Which faculty members are listed in multiple departments for teaching mathematics?
20. Who are the instructors for the B. Tech. and M. Tech. courses?
21. Which faculty members have expertise in multiple disciplines?

## 7. RESULTS AND OUTCOMES

**The final ontology model is as follows:**

**Top-Level Classes:** Thing > Faculty > Department > Specialization > Program

**Middle-Level Classes:** B.Tech, M.Tech, B.Sc, M.Sc, and Research domains

**Bottom-Level Classes:** Specific subjects (e.g., Artificial Intelligence, Organic Chemistry) and named individuals (faculty profiles)

### 7.1 Structural Considerations

The hierarchical design of the ontology offers flexible navigation across departments, programs, and subjects. Object properties, hasExpertiseIn, belongsToDepartment, and teachesInProgram, enhance this structure. Academic libraries can utilize this ontology to provide expert recommendations, develop collections tailored to specific subjects, and support interdisciplinary initiatives. The semantic framework enables the automation of tasks, such as targeted faculty outreach or course planning, based on existing expertise. This structured framework also enables sophisticated enquiries with the exact retrieval of faculty members by multiple attributes. As a dynamic tool, the ontology enables ongoing refinement of academic services to the institution's objectives.(Ochs et al., 2016)

### 7.2 Limitation

The implementation of this model with an institutional database or portal for expert finding would necessitate interfacing with IT departments. In addition, setting up data validation workflows and access by role will be essential to ensure the relevance and integrity of the ontology. Automation processes, such as scheduling data synchronisation and change notifications, can facilitate routine maintenance with minimal hands-on intervention. The training of administrative and library staff in the use and maintenance of the ontology will ensure its long-term sustainability. The final key to effective implementation is institutional buy-in and cross-sectoral cooperation. Although the ontology is scalable, its practical application requires regular updates to keep track of faculty moves, new staff and changing fields of research. (Utkucu & Sacks, 2025b)

## 8. PROPOSED FUTURE WORK

Despite the basis established by this survey to portray the expertise of professors in a university setting using ontological methods, aspects are still awaiting development. Though the outlined ontology shows how academic research services can be enriched by formal knowledge representation, the fullest potential lies in actual deployment and scaling in practice. Future directions should focus on integrating the ontology in institutional databases and digital portals like faculty directories, learning management systems, and library catalogues for seamless expert information access. Timely synchronisation of faculty profiles can be achieved by automating data updates via application programming interfaces (API) interfaced with Human Resources or academic systems. Building intuitive dashboards for visualisation can make the ontology accessible to non-computing users like librarians, students, and administrators, enhancing its usefulness. Expanding the model with research outputs, grant activities, and co-author networks will further enhance the ontology so faculty strengths can be investigated more profoundly and interdisciplinary action encouraged across the institution.(Utkucu & Sacks, 2025a)

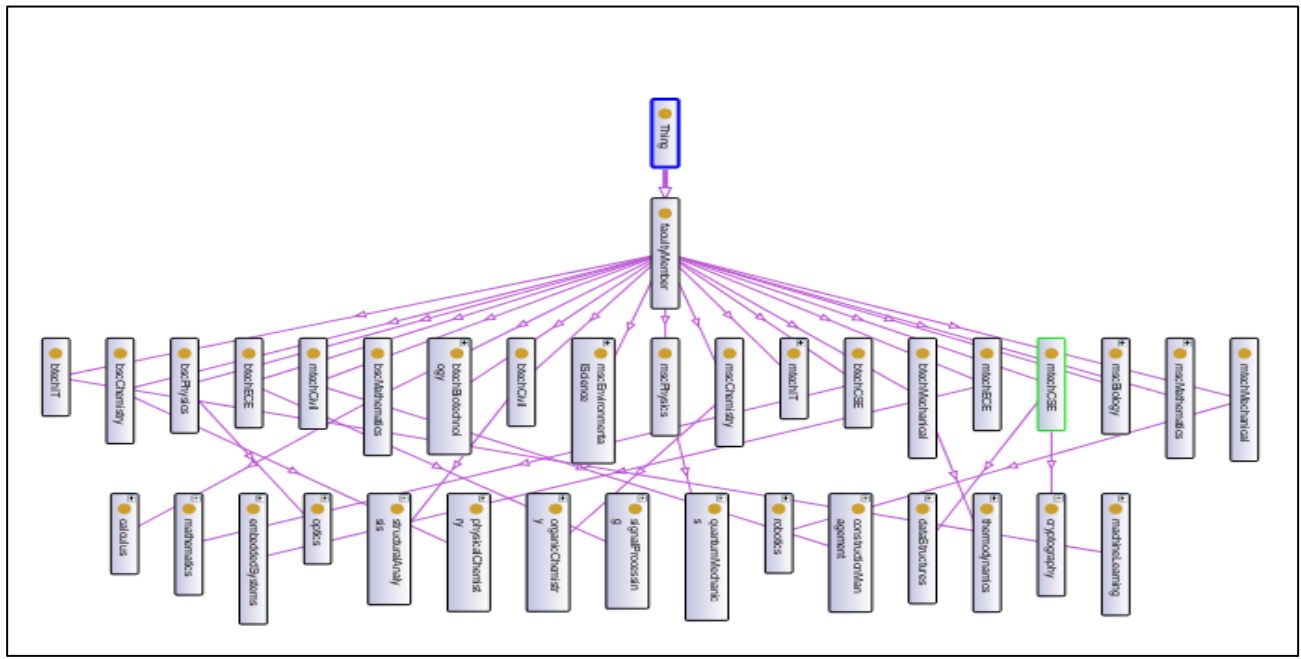

Figure 5: Panoramic view showing disciplinary interconnections

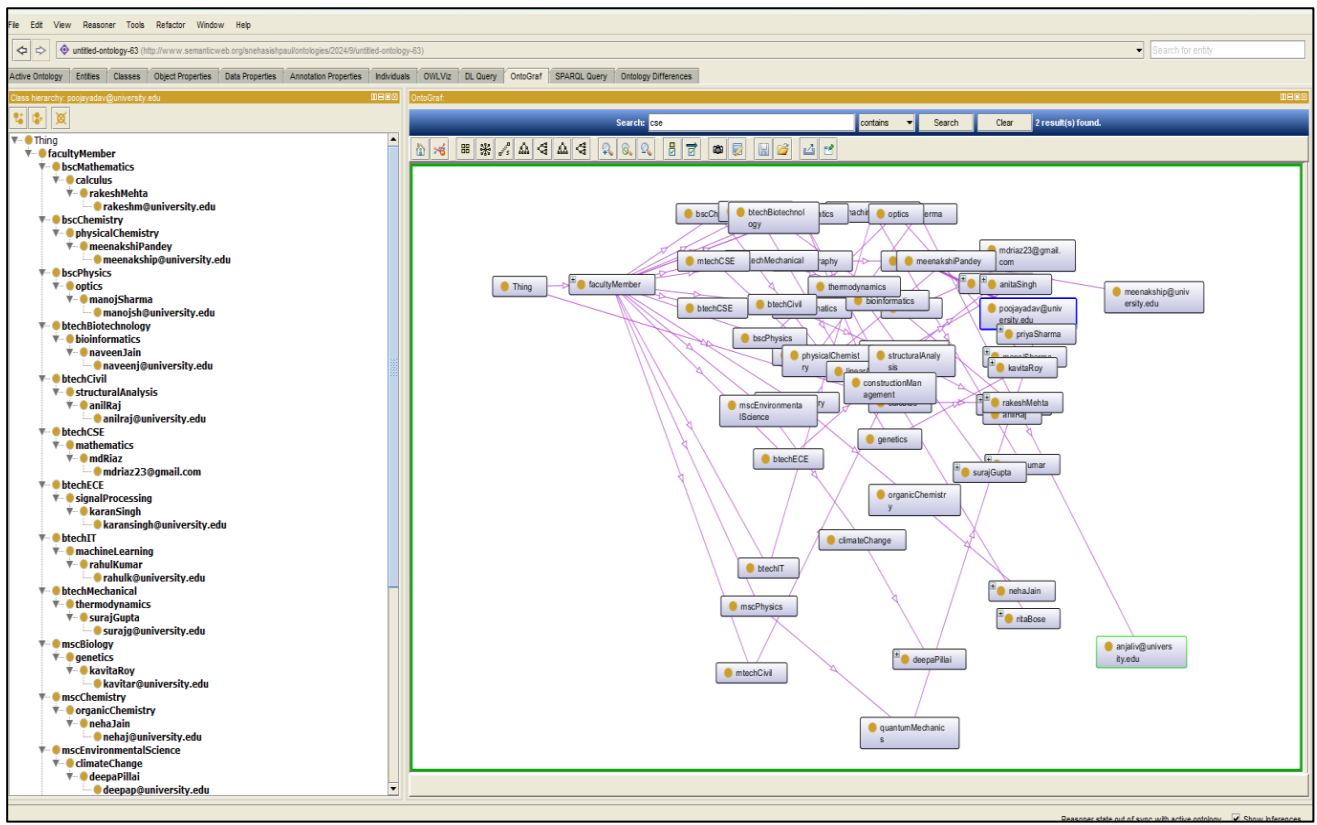

Figure 6: Complete ontology view showing the class hierarchy panel and relationship visualization.

## 9. CONCLUSIONS

The Protégé-based faculty expertise ontology is a landmark in knowledge management in academia. By hierarchically structuring expertise across disciplines, programs, and areas of specialisation with dense semantics, this model changes how institutions assess and leverage internal knowledge resources. Ontology helps librarians to offer referrals to experts and researchers to find collaborators and address knowledge gaps in the institution. Although this semantic structure requires constant updating to cope with organisational changes, it bridges information silos and opens avenues for interdisciplinary collaboration. As institutions increasingly innovate across disciplines, this ontology is not only a valuable operational tool, but also a conceptual framework for redesigning academic knowledge networks along the lines of the semantic web. (Fraga et al., 2020) By combining this ontology with academic databases and institutional repositories, its utility can be magnified exponentially. It forms the foundation for academic analytics powered by AI, intelligent research recommendations and automated expertise mapping. Such developments not only help raise an institution's visibility but also aid in planning and policy development. Finally, the Faculty Expertise Ontology responds to the gap between decision-making and data in higher education. (Hagedorn et al., 2025)

**ACKNOWLEDGEMENT**
I would like to express my deep appreciation to Dr. Mohit Garg for his invaluable guidance and insightful teaching on the development of the ontology. His expertise and mentoring during the course on ontology and semantic web as part of the postgraduate diploma in digital library and data management at the Indira Gandhi National Centre for the Arts (IGNCA) have greatly enhanced my knowledge and skills in this field. I am deeply grateful for his dedication and the clarity with which he imparted complex concepts, which played a crucial role in shaping the foundation of this research.

**Declaration of Conflicting Interests**
The author declares no potential conflicts of interest with respect to the research, authorship, and/or publications of this article.

**Funding**
The author received no financial support for the research, authorship, and/or publication of this article.

**Use of Artificial Intelligence (AI)- Assisted Technology for Manuscript Preparation**
The author Confirm that no AI- Assisted technologies were used in the preparation or writing of the manuscript, and no images were altered using AI

**ORCID**
Snehasish Paul https://orcid.org/0009-0003-2730-5314